\documentclass[a4paper,11pt]{article}

\usepackage{import}
\usepackage{preamble}

\include{preamble.sty}

\title{\boldmath Charge asymmetry in decays $ B\rightarrow D\bar DK$}

\author[]{A.E. Bondar}
\author[]{A.I. Milstein}

\affiliation[]{Budker Institute of Nuclear Physics of SB RAS, 630090 Novosibirsk, Russia}
\affiliation[]{Novosibirsk State University, 630090 Novosibirsk, Russia}

\emailAdd{A.E.Bondar@inp.nsk.su}
\emailAdd{A.I.Milstein@inp.nsk.su}

\abstract{
We discusses the charge asymmetry in $ B\rightarrow D\bar DK$ decays with an invariant mass of the $D\bar D$ pair near the $\Psi(3770)$ resonance. Unlike $ \Psi (3770) $ decays in $ e^+e^- $ annihilation, in $B^+$ decays the probability of $D^0\bar D^0$ production is almost three times higher than $D^+ D^-$. In $B^0$ decays, the ratio of these probabilities  will be opposite. The effect is explained by the fact that, in $B$ -meson decays, the $D\bar D$ pair is produced in a superposition of isoscalar and isovector states, and only in combination with $K$ -mesons  the total state has $ 1/2 $ isospin. We present a simple model  in which the interference of the nonresonant isovector amplitude with the resonant isoscalar amplitude explains the experimental data.
}

\begin{document}
\maketitle
\flushbottom

\section{Introduction}

\begin{figure}[h]
    \centering
    \includegraphics[width=0.7\textwidth]{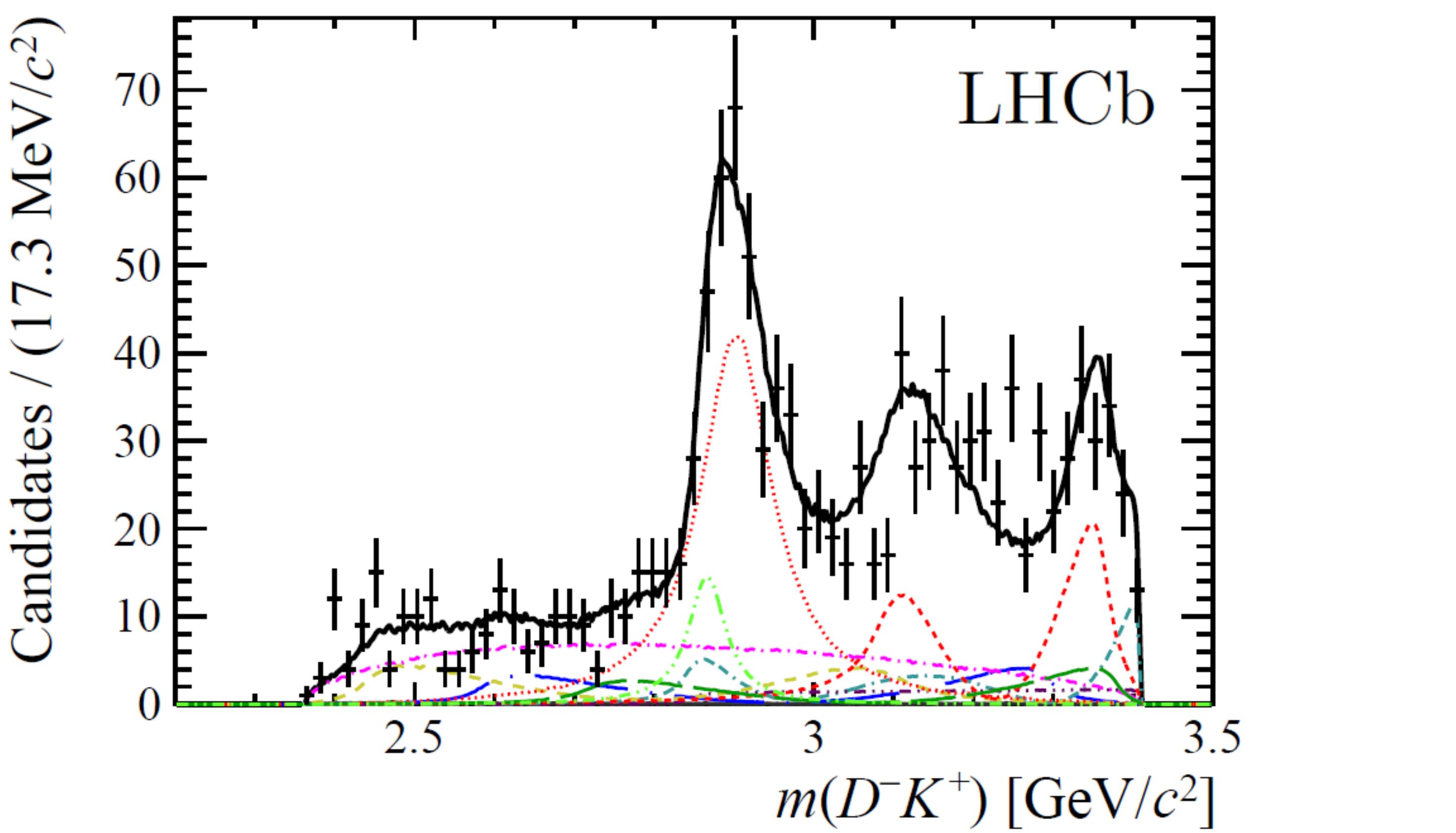}
    \caption{Distribution over the invariant mass of $ D^-K^+ $ in the decay $ B^+ \rightarrow
    	 D^+D^-K^+ $ in the LHCb~\cite {CERN_Seminar} data. The dots show the data, the curves show the resulting fit function and the contributions of the individual components of the model.}
    \label{fig:LHCb_MDK}
\end{figure}

Recently, at LHC seminar at  CERN~\cite{CERN_Seminar,LHCb}, the LHCb collaboration presented preliminary results of amplitude analysis of the decay  $B\rightarrow D^+ D^-K^+$. General attention was drawn to the presence of a peak at an energy 2.9~GeV in the distribution over the invariant mass of $D^-K^+$, Fig.~\ref{fig:LHCb_MDK}. In the short time since the presentation, many articles have appeared offering different interpretations of this phenomenon~\cite{karliner,Mei,Xiao,Xiao1,Jian,Qi,Ming,Hua,Jun,Zhi,Yin}.

\begin{figure}[h]
    \centering
    \includegraphics[width=0.7\textwidth]{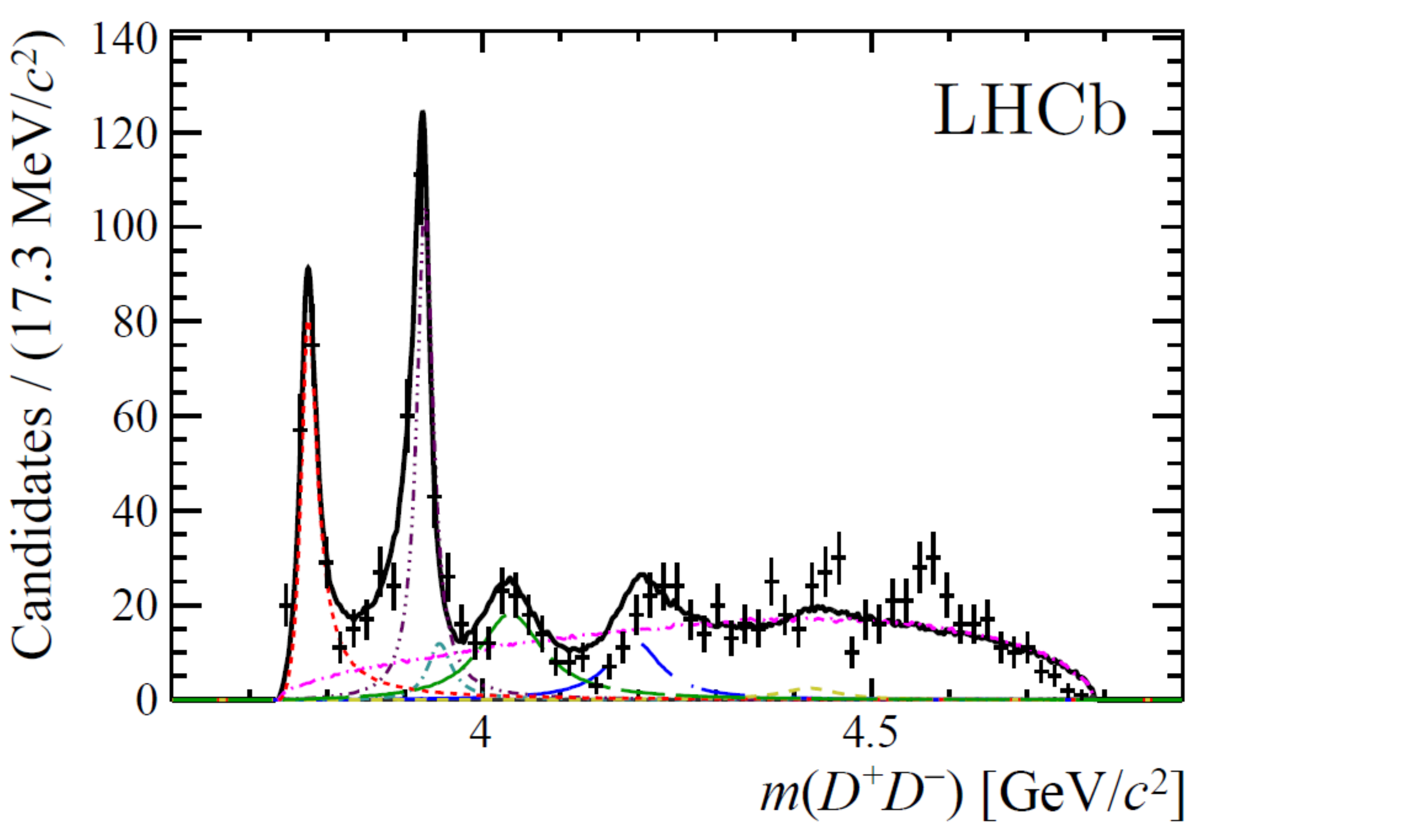}
    \caption{The distribution over the invariant mass of $ D^-D^+ $ in the decay $ B^+ \rightarrow
    	D^+ D^-K^+ $ in the LHCb~\cite {CERN_Seminar} data. The dots show the data, the curves show the resulting fit function and the contributions of the individual components of the model.}
    \label{fig:LHCb_MDD}
\end{figure}

\begin{figure}[h]
	\centering
	\includegraphics[width=0.7\textwidth]{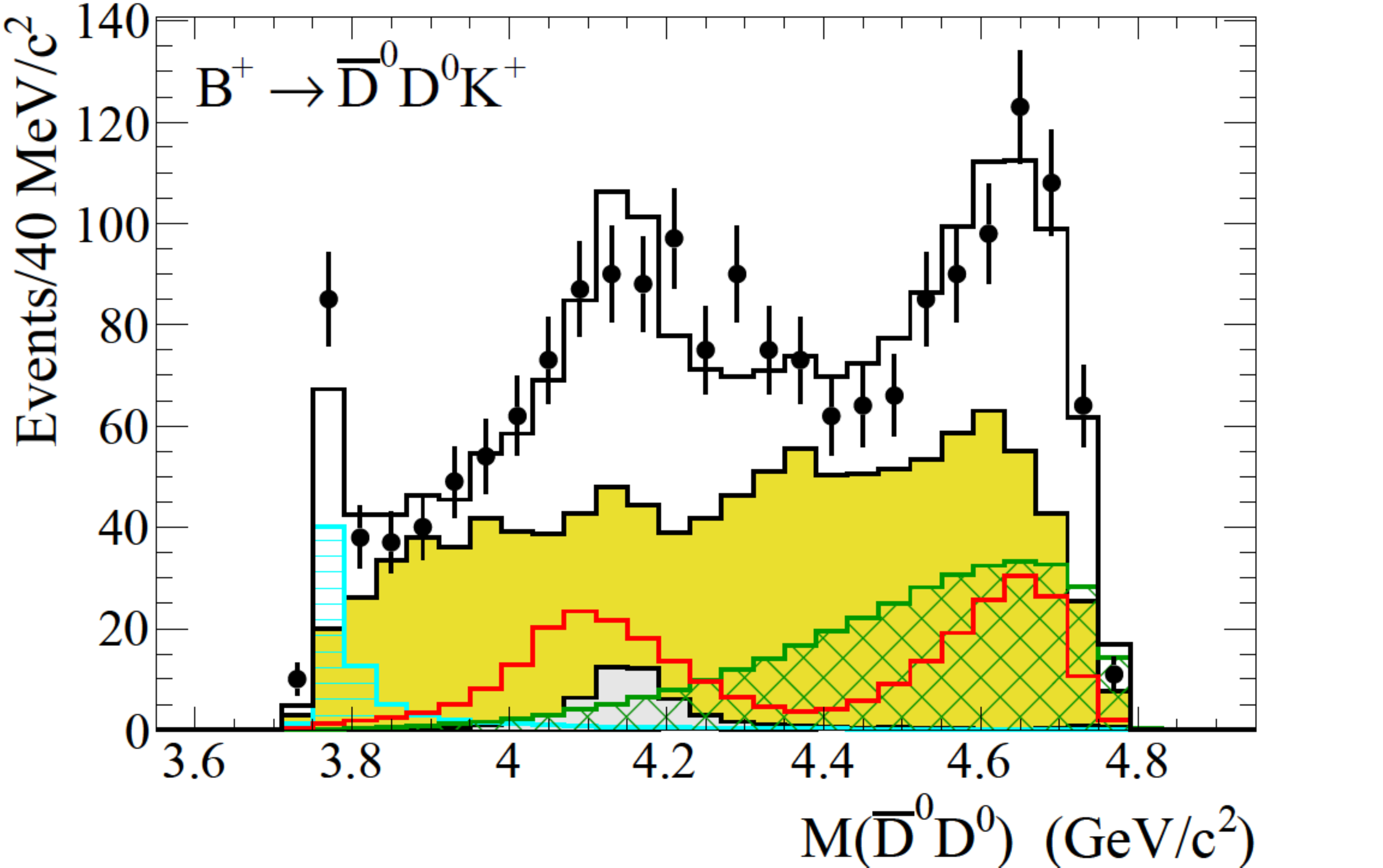}
	\caption{The distribution over the invariant mass of $D^0\bar D^0$  in the decay $B^+ \rightarrow D^0\bar D^0K^+$ in the Babar~\cite{Babar} data. The dots show the data, the histograms show the resulting description by the model and individual contributions.}
	\label{fig:Babar_MD0D0}
\end{figure}

These interpretations are based on the  hypotheses on the production of a compact $\bar c\bar s ud$ tetraquark, $D^*K^*$ molecules, etc. However, no one paid attention to another interesting phenomenon that is clearly manifested in the LHCb data. In the distribution over the invariant mass $D^+D^-$ (Fig.~\ref{fig:LHCb_MDD}) in the decay $B^+ \rightarrow D^+D^-K^+$, two peaks are observed, which are interpreted by the authors~\cite{CERN_Seminar}, as signals of charmonia $\Psi(3770) $, $\chi_{c0} (3930)$, and $\chi_{c2} (3930)$. It would seem that such an interpretation is natural. However, if we look at the invariant mass distribution of $D^0\bar D^0$ in the decay $B^+\rightarrow D^0\bar  D^0K^+$~\cite{Babar}, Fig.~\ref{fig:Babar_MD0D0}, then we will see only peak $\Psi(3770)$.
At first glance, we observe a contradiction, since the isotopic spin of charmonia is zero, and, therefore, the probabilities of their decays into $D^+D^-$  and $ D^0\bar D^0$  should be equal. This is precisely what is observed in the decays of $\Psi(3770)$  produced in $e^+e^-$ annihilation. Our work is devoted to the possible interpretation of the apparent contradiction.

\section{\boldmath Charge asymmetry in $B\to D\bar{D} K$ decays}
Assume that the masses of charged and neutral $D$-mesons coincide, i.e., the violation of isotopic invariance associated with the difference  of   $u$ and $d $ quark   masses is absent. Consider the production of $D\bar  D$ pairs  in $B^+$ decays with an invariant mass near $ M = 3770\,\mbox{ MeV}$ and  estimate in this region the ratio of the decay probabilities
$$R=W(B^+\rightarrow D^+D^-K^+)/W(B^+\rightarrow \bar D^0D^0K^+).$$

\begin{figure}[h]
    \centering
    \includegraphics[width=1.0\textwidth]{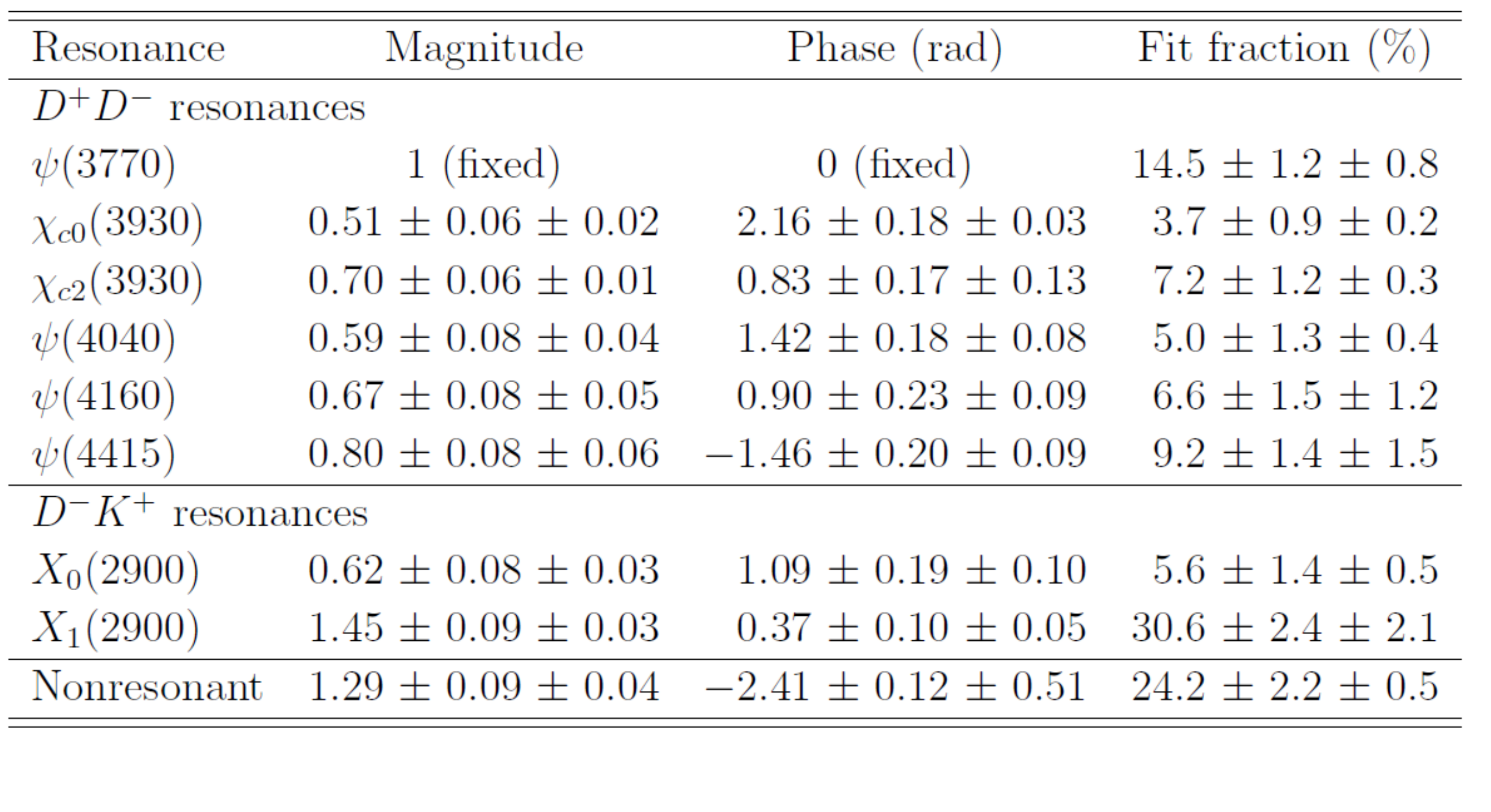}
    \caption{Fitfractions, amplitudes and phases of two-particle intermediate states in the amplitude analysis of the $B^+ \rightarrow D^+D^-K^+$  decay in the LHCb~\cite{CERN_Seminar} data}
    \label{fig:LHCb_Table}
\end{figure}

Since the LHCb~\cite{CERN_Seminar} does not present the  absolute decay probability $W_{tot}^{+-}$ of the decay $B^+\rightarrow D^+ D^-K^+$, but only the fraction $W_{Res}^{+-}/W_{tot}^{+-}$ of this decay probabilities in the vicinity of $\Psi(3770)$ resonance, see Table~\ref{fig:LHCb_Table},  then we  use the  results of Babar~\cite{Babar1},  $W_{tot}^{+-}=(2.2\pm 0.5\pm 0.5)\cdot 10^{-4}$, the result of LHCb (Tab.~\ref{fig:LHCb_Table}), and obtain $W_{Res}^{+-}=(3.2 \pm 0.77 \pm 0.75)\cdot 10^{-5}$. Comparing this value with the corresponding value $W_{Res}^{00}=(11.8 \pm 4.1 \pm 1.5)\cdot 10^{-5}$, for the decay $B^+\rightarrow  D^0\bar D^0K^+$ \cite{Babar}, 
we find the ratio $R=0.27\pm 0.13$ (the statistical and systematic errors were added quadratically to the total uncertainty), which is three times less than the ratio obtained in $e^+e^-$ annihilation by CLEO~\cite{CLEO}, $0.799 \pm 0.006 \pm 0.008$, and by BESIII~\cite{BESIII}, $0.7823 \pm 0.0036 \pm 0.0093$.

The $R$ ratio can also be obtained from the Belle~\cite{Belle} and Babar~\cite{Babar2} data. 
Note that both measurements did not use amplitude analysis to obtain the number of events.
Belle~\cite{Belle} obtained the ratio $R=0.41 \pm 0.25 \pm 0.073$. Babar~\cite{Babar2} does not present the corresponding value, although it follows from the data of \cite{Babar2}  that $R=0.6 \pm 0.31$. It is seen that the experimental accuracy of B-factories is insufficient for any unambiguous conclusions. The  ratio $R $ derived from LHCb data is more accurate than previously published values
and is consistent with them within errors. Thus, we see that  the signal $D^0\bar D^0$ 
in the resonance region  is many times greater than that  of $D^+D^-$. Perhaps this  effect explains the apparent difference in the probability of $D^0\bar D^0$ production  in the region of invariant masses of $\Psi(3770)$ and $\chi_{c0} / \chi_{c2}$. In addition, we come to the important conclusion that the hadronic system of $D\bar D$ produced in $B^+$ decay in the vicinity of $\Psi (3770)$  differs from the resonance $\Psi (3770)$ observed in $e^+e^-$ annihilation. How can this large charge asymmetry be explained?

In terms of quarks, the decay $B^+\rightarrow D\bar DK$ corresponds to the process
$$u\,\bar b\rightarrow u\,(c\,\bar c\,\bar s)(u\,\bar u+d\,\bar d),$$
where we took into account a light quark-antiquark pair  with zero isospin ($u\,\bar u+d\,\bar d$) produced from the vacuum. There are two options: the spectator $u$  goes into a bound state with the antiquark $\bar s$  or with $\bar c$.  As a result, the wave function $\psi$  of the final state can be written  as
$$\psi=a\,\dfrac{(D^0\,\bar D^0-D^+\, D^-)}{\sqrt{2}}\,K^++b\,\frac{(D^0\, K^+-D^+\,K^0)}{\sqrt{2}}\,\bar D^0\,.$$
Here the states in parentheses have an isospin equal to zero; we consider only the quark composition and not discuss  spin, angular momentum or other quantum numbers. The wave function $\psi$ must be rewritten in terms of quasiparticles, which are systems of strongly interacting $D\bar D$. From the isospin point of view, there are two such systems: states with isospin zero and one. We have
 \begin{align}
 &\psi=(a+b/2)\psi_0+\dfrac{\sqrt{3}}{2}\,b\,\psi_1\,,\nonumber\\
&\psi_0=|0,0\rangle\,  |1/2,1/2\rangle\nonumber\\
&=\dfrac{(D^0\,\bar D^0-D^+\, D^-)}{\sqrt{2}}\,K^+\nonumber\\
 &\psi_1=\dfrac{1}{\sqrt{3}}|1,0\rangle\,  |1/2,1/2\rangle-\sqrt{\dfrac{2}{3}}|1,1\rangle\,  |1/2,-1/2\rangle\nonumber\\
&= \dfrac{1}{\sqrt{3}}\,\frac{(D^0\,\bar D^0\,+D^+\,\bar D^-) }{\sqrt{2}}\,K^+-\sqrt{\dfrac{2}{3}}D^+\,\bar D^0\,K^0\,.
\end{align}
Thus, $\psi_0$  is a system consisting of interacting $D\bar D$  with isospin zero and $K^+$, $\psi_1$   is a system consisting of interacting $D\bar D$  with isospin one and $K^+$  or $K^0$ with total isospin $ 1/2 $ and projection $ + 1/2 $. The problem under discussion is similar to the well-known problem of the production of nucleon-antinucleon pairs in $e^+e^-$  annihilation near the threshold, since in this case the  hadronic state is a superposition of the isovector and isoscalar parts. The non-trivial dependence of the cross section on the energy near the pair production threshold is explained by the interaction of slow  nucleons and antinucleons through a strong optical potential \cite{DMS2014, DMS2016, MS2018}. Optical potentials are different for the isovector and isoscalar states. The imaginary part of the optical potential takes into account the processes of annihilation of a nucleon-antinucleon pair into mesons.

Since our goal is not to obtain accurate predictions, which is a very non-trivial task, but to explain the phenomenon at the qualitative level, we  consider the simplest model, which, nevertheless, contains all essential features of a real problem.

Consider the simplest case $M_c=M_0=M$,  where $M_c$  and $M_0$  are masses, respectively, of charged and neutral $D$-mesons.
The optical interaction potential  is denoted by $U_0(r)$  for the isosinglet state and by $U_1(r)$ for the isotriplet state. For simplicity, we   assume that the $D\bar D$ pair is in a state with an orbital angular momentum $l=0$ (as will be explained below, the asymmetry mechanism for the case $l=1$  does not qualitatively differ from the case $l=0$). To find the decay probability, we use the approach described in the works \cite{DMS2014, DMS2016, MS2018}. First, it is necessary to find regular solutions  $ u_n (r) $ of the equations
$$\left[\dfrac{p_r^2}{M} +U_n(r)-E\right]u_n(r)=0\,,\quad n=0,\,1\,, $$
where $(-p^2_r)$  is the radial part of the Laplace operator. For $r\rightarrow \infty$, the asymptotic form of the solutions is
\begin{align}
&u_n(r)=\dfrac{1}{2i}[S_n\,\chi_k^{+}-\chi_k^-]\,,\quad |S_n|\leq 1\,,\nonumber\\
&\chi_k^+=\dfrac{\exp(ikr) }{ kr}\,,\quad \chi_k^-=\dfrac{ \exp(-ikr)}{ kr}\,,\quad k=\sqrt{ME}\,.
\end{align}  
After that, the probabilities   $W^{+-}$, $W^{00}$, 
 and $W^{+0}$   of decays, respectively, $B^+\rightarrow D^+D^-K^+$, $B^+\rightarrow D^0\bar D^0K^+$  and $B^+\rightarrow D^+\bar D^0K^0$ can be expressed up to a common factor in terms of $u_n(0)$  as follows 
\begin{align}\label{3W}
&W^{+-}=\left|-\left(a+\dfrac{b}{2} \right)\,u_0(0)+\dfrac{b}{2} \,u_1(0)\right|^2\,,\nonumber\\
&W^{00}=\left|\left(a+\dfrac{b}{2} \right)\,u_0(0)+\dfrac{b}{2} \,u_1(0)\right|^2\,,\nonumber\\
&W^{+0}=| b\,u_1(0)|^2\,.
\end{align}
The charge asymmetry is determined not only by the values of $a$ and $b$, which can be considered energy independent near the  threshold of   $D\bar D$  pair production, but also by the values of the functions $u_0(0)$ and $u_1(0)$ having the energy dependence  determined by the  isoscalar and isovector optical potentials, respectively. To explain the charge asymmetry, it is convenient to introduce the variable
\begin{equation}\label{xx}
x=\left(\dfrac{2a}{b}+1\right)\,\dfrac{u_0(0)}{u_1(0)}\,
\end{equation}
and rewrite the expressions for the probabilities as
\begin{align}\label{3Wnew}
&W^{+-}=\dfrac{1}{4}F\left|x-1\right|^2\,,\quad W^{00}=\dfrac{1}{4}F\left|x+1\right|^2\,,\quad
W^{+0}=F\,,
\end{align}
where $F$  is some function of energy that does not affect the probability ratio. All information about the charge asymmetry is contained in the variable $x$, which is, generally speaking, a complex quantity. 

\begin{figure}[h]
    \centering
    \includegraphics[width=0.7\textwidth]{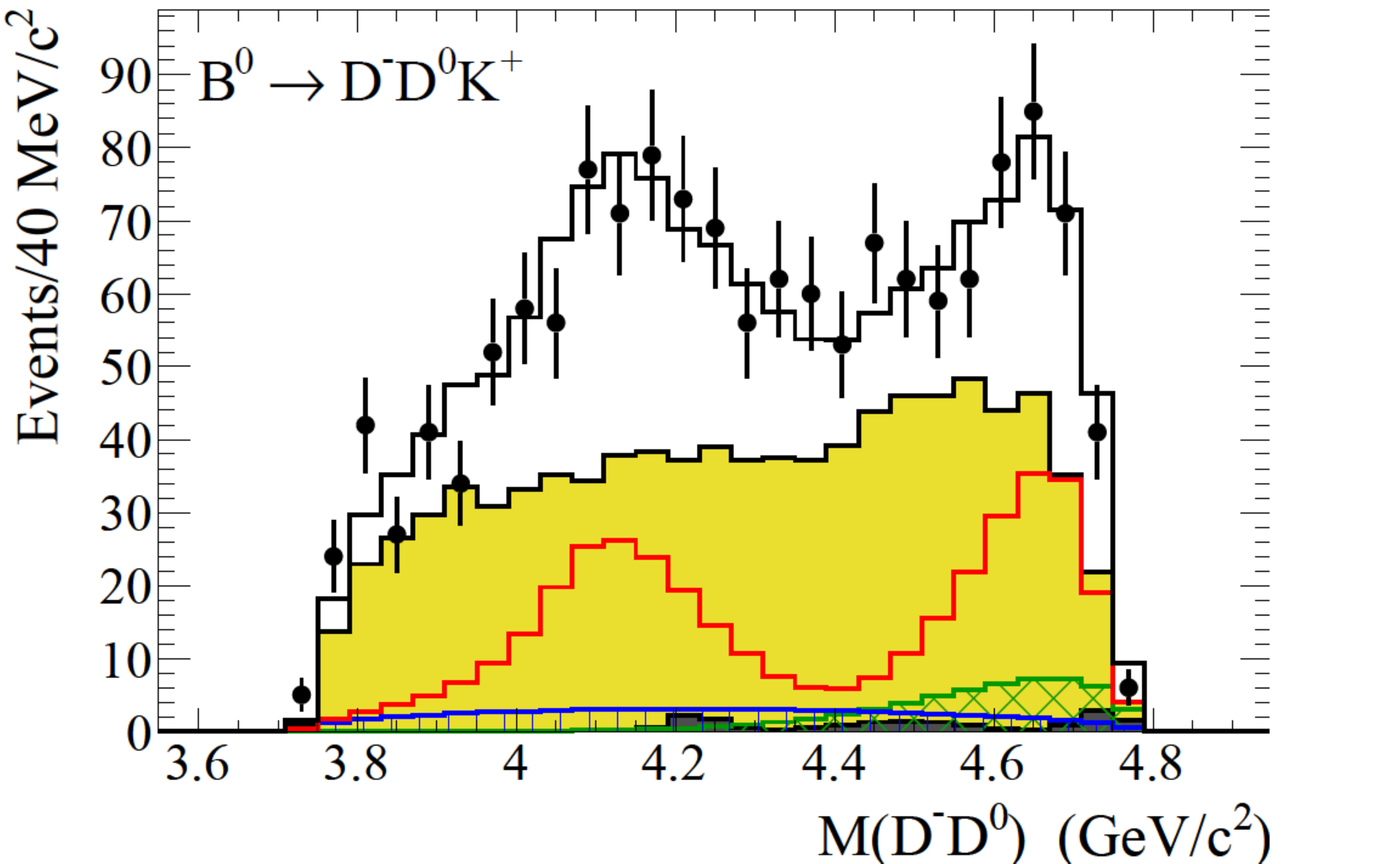}
    \caption{The invariant mass distribution of $  D^0D^-$ in the decay $B^0 \rightarrow   D^0D^-K^+$ in the Babar~\cite{Babar} data. The dots show the data, the histograms show the resulting description by the model and individual contributions.}
    \label{fig:Babar_MD+D0}
\end{figure}

Similar expressions can be obtained for the decay probabilities of a neutral $B$-meson, $B^0 \rightarrow \bar DDK$. In this decay
the wave function  $\widetilde \psi$ of the final state is
\begin{align}
&\widetilde\psi=(a+b/2)\widetilde\psi_0-\dfrac{\sqrt{3}}{2}\,b\,\widetilde\psi_1\,,\nonumber\\
&\widetilde\psi_0=\dfrac{(D^0\,\bar D^0-D^+\, D^-)}{\sqrt{2}}\,K^0\nonumber\\
&\widetilde\psi_1=\dfrac{1}{\sqrt{3}}\,\frac{(D^0\,\bar D^0\,+D^+\,\bar D^-) }{\sqrt{2}}\,K^0-\sqrt{\dfrac{2}{3}}D^0\,  D^-\,K^+\,.
\end{align}
Therefore, for the probabilities $\tilde W^{+-}$, $\tilde W^{00}$ and $\tilde W^{0-}$ in $B^0 $ decay into $D^+D^-K^0$, $D^0\bar D^0K^0$,  and $D^0  D^-K^+$, respectively, we obtain:
\begin{align}\label{123WT}
&\widetilde W^{+-}=W^{00}\,,\quad \widetilde W^{00}=W^{+-}\,,\quad \widetilde W^{0-}=W^{+0}\,.
\end{align}
Since it follows from the experiment that   peaks in the invariant mass of $D^0D^-$  in the energy region of $\Psi(3770)$ resonance are not observed in the decay $ B^0\rightarrow D^0D^-K^+$, see 
Fig.~\ref{fig:Babar_MD+D0}, then it is natural to consider the function $u_1(0)$ to be  independent of energy. The energy dependence of the function $u_0(0)$  has a resonant form and can be found from the cross section of $\Psi(3770)$  production   in $e^+e^-$  annihilation, in which $D\bar D$  is   in the isoscalar state. As a result, the dependence of the function $x$  on energy has the form:
\begin{equation}\label{BW}
x= \dfrac{C}{E-E_0+i\Gamma/2}\,,
\end{equation}
where  $E_0=\Gamma=30\,\mbox{MeV}$, and $C$ is  some complex parameter. 

\begin{figure}[h]
	\centering
	\includegraphics[width=0.45\linewidth]{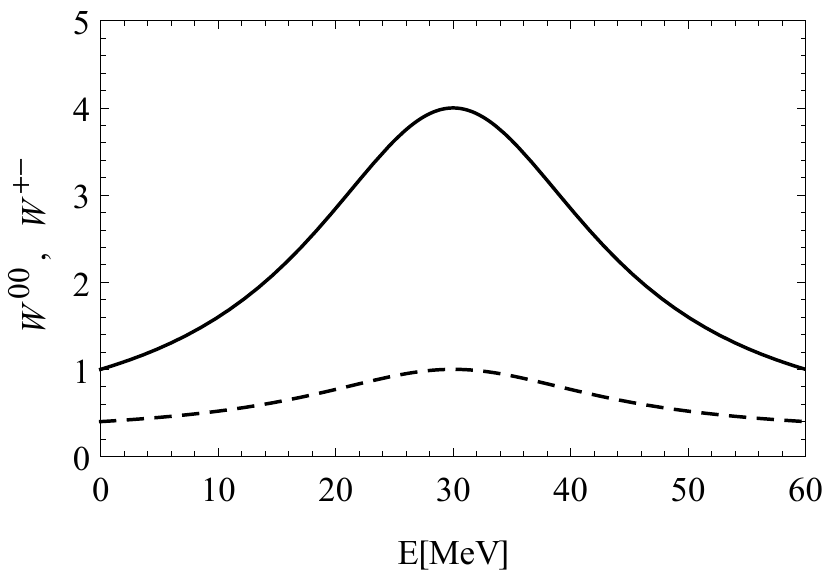}
	\includegraphics[width=0.45\linewidth]{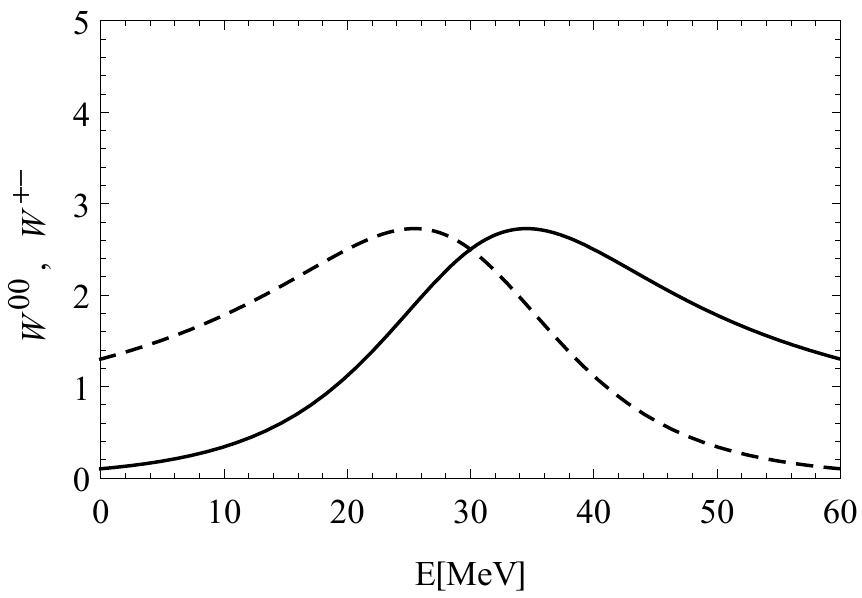}
	\caption{Energy dependence of the probabilities $W^{00}$ (solid line) and $W^{+-}$ (dashed line) in units of $W^{+0}$ for $ | C | = 45\,\mbox{MeV} $, $\arg C=\pi/2$  (left figure) and $\arg C=0$  (right figure). }
	\label{fig:wwfig}
\end{figure}
Fig.~\ref{fig:wwfig} shows the energy dependence of the probabilities $W^{00}$ and $W^{+-}$ in units of $ W^{+0}$ for $|C|=45\,\mbox{MeV}$ , which reproduces the experimentally observed charge asymmetry for some $\arg C$. These dependencies are very sensitive to the phase value of the  parameter $C$. For $\arg C=\pi/2$, good agreement with experiment is seen. The value of charge asymmetry $R$ for the selected parameters is $0.25$. For $\arg C=0$, the charge asymmetry disappears. It is important that the interference between the resonant isoscalar and nonresonant isovector amplitudes can lead not only to charge asymmetry, but also to distortion of the resonance shape and its parameters in different decay channels.

We emphasize that the picture of charge asymmetry described above for the case of the $\bar DD$ system with orbital angular momentum $l=0$  is completely preserved for $l=1$. The only modification is an explicit expression for the function $x$~\eqref{xx}, where the ratio of derivatives $u_0'(0)/u_1'(0)$  should be used instead of the ratio $u_0(0)/u_1(0)$. However, the resonant form of the function $x$~\eqref{BW} is preserved.

\section{Conclusion}

In our work, we indicated a large difference in the  probabilities of $B^+ \rightarrow  D^+D^-K^+$ and $B^+ \rightarrow D^0\bar D^0K^+$ decays for the invariant mass of $D\bar D$ pair in the vicinity of the resonance $\Psi (3770)$. This difference follows  from the experimental data of LHCb and Babar. 
The ratio of the probabilities is $R=0.27 \pm 0.13\,$. It is shown that such a large charge asymmetry is apparently related to the interference between the resonant isoscalar and nonresonant isovector amplitudes of $D\bar D$ pair production. The simple model we constructed is in good agreement with the experimental data. We predict that, up to the effects associated with a small difference in the masses of charged and neutral $ D $-mesons, the value of $ R $ will be inverse in the decay of $B^0$ meson. Similar effects should be expected in other decays of $ B $ -mesons, such as $B \rightarrow D^{(*)}\bar D^{(*)}K$, for invariant masses of $D^{(*)}\bar D^{(*)}$ near the corresponding resonances. We have shown that the interference between the resonant isoscalar and nonresonant isovector amplitudes can lead not only to charge asymmetry, but also to a significant distortion of the resonance shape and its parameters in different decay channels.
Therefore, it is important to take into account the contribution of the isovector amplitude in the amplitude analysis of $ B $-meson decays.
The relations~\eqref{3W} allow us to formulate a recipe for the correct extraction of the $D\bar D$ isoscalar resonance contribution  to the decay probability $B \rightarrow D\bar DK$. This contribution is determined by the combination $W^{+-}+W^{00}-W^{+0}/2$. Namely,   the Dalitz plot corresponding to the isoscalar contribution to the decay probability of the $B$ meson can be obtained by combining the densities of events corresponding to each channel, according to the combination pointed out above.

We are grateful to Anton Poluektov for valuable discussions.



\begin{thebibliography}{9}
\bibitem{CERN_Seminar}
\textit{LHC Seminar, $B\rightarrow D\bar DK$ decays: A new (virtual) laboratory for exotic particle searches at LHCb, by Danial Johnson, CERN, August 11, 2020,}
{https://indico.cern.ch/event/900975/}

\bibitem{LHCb}
LHCb Collaboration, Roel Aaij(NIKHEF, Amsterdam) et al. ,
\textit{"Amplitude analysis of the $B^+ \to D^+D^-K^+$ decay"},

{e-Print: 2009.00026 [hep-ph]}.

\bibitem{karliner}
M.~Karliner and J.~Rosner,
\textit{"First exotic hadron with open heavy flavor: $cs\bar u\bar d$ tetraquark"},

{e-Print: 2008.05993 [hep-ph]}.


\bibitem{Mei}
Mei-Wei Hu,Xue-Yi Lao,Pan Ling,Qian Wang,
\textit{"The $X_0(2900)$ and its heavy quark spin partners in molecular picture"},

{e-Print: 2008.06894 [hep-ph]}.


\bibitem{Xiao}
Xiao-Gang He,Wei Wang,Ruilin Zhu,
\textit{"Open-charm tetraquark $X_c$ and open-bottom tetraquark $X_b$"},

{e-Print: 2008.07145 [hep-ph]}.

\bibitem{Xiao1}
Xiao-Hai Liu,Mao-Jun Yan,Hong-Wei Ke,Gang Li,Ju-Jun Xie,
\textit{"Triangle singularity as the origin of $X_0(2900)$ and $X_1(2900)$ observed in $B^+\to D^+ D^- K^+$"},

{e-Print: 2008.07190 [hep-ph]}.

\bibitem{Jian}
Jian-Rong Zhang,
\textit{"An open charm tetraquark candidate: note on $X_{0}(2900)$"},

{e-Print: 2008.07295 [hep-ph]}.

\bibitem{Qi}
Qi-Fang Lü,Dian-Yong Chen,Yu-Bing Dong,
\textit{"Open charm and bottom tetraquarks in an extended relativized quark model"},

{e-Print: 2008.07340 [hep-ph]}.



\bibitem{Ming}
Ming-Zhu Liu,Ju-Jun Xie,Li-Sheng Geng,
\textit{"$X_0(2866)$ as a $D^*\bar{K}$ molecular state"},

{e-Print: 2008.07389 [hep-ph]}.

\bibitem{Hua}
Hua-Xing Chen,Wei Chen,Rui-Rui Dong,Niu Su,
\textit{"$X_0(2900)$ and $X_1(2900)$: hadronic molecules or compact tetraquarks"},

{e-Print: 2008.07516 [hep-ph]}.


\bibitem{Jun}
Jun He,Dian-Yong Chen,
\textit{"Molecular picture for $X_{0}(2900)$ and $X_1(2900)$"},

{e-Print: 2008.07782 [hep-ph]}.

\bibitem{Zhi}
Zhi-Gang Wang,
\textit{"Analysis of the $X_0(2900)$ as the scalar tetraquark state via the QCD sum rules"},

{e-Print: 2008.07833 [hep-ph]}.

\bibitem{Yin}
Yin Huang,Jun-Xu Lu,Ju-Jun Xie,Li-Sheng Geng,
\textit{"Strong decays of $\bar{D}^{*}K^{*}$ 
molecules and the newly observed $X_{0,1}$ states"},

{e-Print: 2008.07959 [hep-ph]}.


\bibitem{Babar}
BaBar Collaboration, J.P.Lees (Annecy, LAPP) \textit{et al}, 
\textit{"Dalitz plot analyses of $B^0 \rightarrow D^- D^0 K^+$ and $ B^+ \rightarrow \bar D^0D^0K^+$ decays"},
{Phys. Rev. \textbf{D91} (2015) 052002}.
\bibitem{Babar1}
BaBar Collaboration, P. del Amo Sanchez(Annecy, LAPP) \textit{et al}, 
\textit{"Measurement of the $B\rightarrow \bar D^{(*)}D^{(*)}K$ branching fractions"},
{Phys. Rev. \textbf{D83} (2011) 032004}.

\bibitem{CLEO}
CLEO Collaboration, G.Bonvicini(Wayne State U.) \textit{et al}, 
\textit{"Updated measurements of absolute $D^+$ and $D^0$ hadronic branching fractions and $\sigma (e^+e^- \rightarrow \bar DD)$ at $E_{cm}=3774$ MeV"}
Phys. Rev. D89 7, (2014) 072002.

\bibitem{BESIII}
BESIII Collaboration, Medina Ablikim (Beijing, Inst. High Energy Phys.) \textit{et al}, 
\textit{"Measurement of $e^+e^- \rightarrow \bar DDK$ cross section at the $\Psi (3770)$ resonance"},
{Chin. Phys.  \textbf{C42} (2018) 8, 083001}.

\bibitem{Babar2}
BaBar Collaboration, Bernard Aubert (Annecy, LAPP) \textit{et al}, 
\textit{"Study of Resonances in Exclusive B Decays to $\bar D^{(*)} D^{(*)}$"},
{Phys. Rev. \textbf{D77} (2008) 011102}.


\bibitem{Belle}
Belle Collaboration, Kazuo Abe(KEK, Tsukuba) \textit{et al}, 
\textit{"Observation of $B^+ \rightarrow \Psi (3770) K^+$"},
{Phys. Rev. Lett. \textbf{93} (2004) 051803}.

\bibitem{DMS2014}  V. F. Dmitriev, A. I. Milstein, S. G. Salnikov,
``Isoscalar amplitude dominance in e+e- annihilation to $N\bar N$ pair close to the threshold'',
Yad. Fiz. {\bf 77}, 1234 (2014) [Physics of Atomic Nuclei {\bf 77},  1173 (2014)].
	
\bibitem{DMS2016}  V.F. Dmitriev, A.I. Milstein, and S.G. Salnikov,
	``Real and virtual $N\bar N$ pair production near the threshold'',
	Phys. Rev. {\bf D 93}, 034033 (2016).
	
\bibitem{MS2018}  A.I. Milstein,  S.G. Salnikov,
	``Fine structure of the cross sections of $e^+e^-$ annihilation
	near the thresholds of $p \bar{p}$  and $n\bar n$ production'',
	Nuclear Physics {\bf A 977}, 60  (2018).
	
\end{thebibliography}
 \end{document}